\documentclass[
    ,final            
  ]
  {aipproc}

\layoutstyle{6x9}

\def\H{H\hskip-8.5pt/\hskip2pt}

\def\lsim{\mathrel{\mathpalette\@versim<}}

\def\gsim{\mathrel{\mathpalette\@versim>}}

\begin{document}

\title{CPT Violation: What and where to look for}

\classification{
                \texttt{11.30.Er  04.70.Dy 03.65.Ud  03.65.Yz  13.25.Es  
14.60.Pq }}
\keywords      {CPT, Quantum Gravity, Decoherence, Phenomenology}

\author{Nick E. Mavromatos}{
  address={King's College London, Department of Physics, London WC2R 2LS, U.K.}
}


\begin{abstract}
In this review I classify the possible ways of CPT violation, and 
I describe briefly their phenomenology, in both terrestrial and astrophysical
experiments, including antimatter factories, neutral mesons 
and neutrinos, and discuss the various sensitivities.
I also pay attention to disentangling genuine quantum-gravity
induced CPT violation from `fake' violation due to ordinary matter
effects. A particularly interesting situation
arises when the breaking of CPT invariance is through unitarity 
violations, in the sense
of the matter theory being viewed as an effective field theory, entangled
with decoherening quantum gravity ``environments''. In such a case the quantum 
mechanical CPT operator  is {\it ill defined} due to another 
mathematical theorem, and 
one has novel effects 
associated with CPT Violating modifications of Einstein-Podolsky-Rosen  
type correlations of entangled meson states in 
B and $\phi$ meson factories.

\end{abstract}

\maketitle


\section{Introduction: CPT Theorem} 
A century after Einstein's {\it annus mirabilis}, and ninety years
after his revolutionary proposal on the dynamical nature of space time,
which is the basis for the classical theory of General Relativity, 
we are still lacking a consistent theory that would describe 
the quantum nature of space time at short-distance 
scales, of order of the Planck 
length $10^{-35}$ m. 
Any complete theory of quantum gravity is bound to address
fundamental issues, directly related to the 
emergence of space-time and its structure at energies 
beyond the Planck energy scale $M_P \sim 10^{19} $ GeV.  
From our relatively low energy experience so far, we are lead to expect
that a theory of quantum gravity should respect most of the fundamental 
symmetries of particle physics, 
that govern the standard model of electroweak and strong 
interactions: Lorentz symmetry and CPT invariance,
that is invariance under the combined action of Charge Conjugation (C),
Parity (reflection P) and Time Reversal Symmetry (T).
Actually the latter invariance is a theorem of any local quantum field theory
that we can use to describe the standard phenomenology of particle
physics to date. The {\it CPT theorem} can be stated as follows~\cite{cpt}: 
{\it Any quantum theory, formulated on {\it flat space time} is symmetric
under the combined action of CPT transformations, provided the theory  
respects} (i) {\it Locality}, (ii) {\it Unitarity} (i.e. conservation of 
probability) and (iii) 
{\it Lorentz invariance}.

\section{Potential CPT Violation}

If such a theorem exists, then why do we have to bother to test
CPT invariance, given that all our phenomenology up to now has been 
based on such quantum theories ? The answer to this question is 
intimately linked with our understanding of {\it quantum gravity}.

\subsection{CPT Violation through Decoherence: Quantum Gravity as a `medium',
opening up the matter quantum system}

First of all, the theorem is not valid (at least in its strong form) 
in highly curved ({\it singular}) 
space times,
such as black holes, or in general in space-time 
backgrounds of some quantum gravity theories 
involving the so-called {\it quantum 
space-time foam} backgrounds~\cite{wheeler}, 
that is {\it singular} quantum fluctuations of space time geometry,
such as black holes {\it etc}, with event horizons 
of microscopic Planckian size ($10^{-35}$ meters). 
Such backgrounds result in {\it apparent} violations of {\it unitarity}
in the following sense: there is part of 
information (quantum numbers of incoming matter) 
``disappearing'' inside the  microscopic event 
horizons, so that an observer at asymptotic infinity will have to 
trace over such ``trapped'' degrees of freedom. 
This would define a ``space time foam'' situation~\cite{wheeler}, 
in which the quantum gravity ground state resembles that of a 
decoherening ``medium'' in open system quantum 
mechanics~\cite{ehns,lopez,mavromatosdecohe}. 
Barring the exciting possibility of holographic properties of 
such microscopic black holes~\cite{maldacena}, 
one may face, under such circumstances, a 
situation in which an initially pure state evolves in time
to get mixed: the asymptotic states are described by density matrices,
defined as follows: $ \rho _{\rm out} = {\rm Tr}_{M} |\psi ><\psi|$, 
where the trace is over trapped (unobserved) quantum states, that disappeared 
inside the microscopic event horizons in the foam. 
Such a non-unitary evolution results in the impossibility of defining 
a standard quantum-mechanical scattering matrix, connecting asymptotic
states in a scattering process:
$|{\rm out}> = S~|{\rm in}>$, $S=e^{iH(t_i - t_f)}$, 
where $t_i - t_f$ is the duration of the scattering (assumed much longer than
other time scales in the problem). Instead, in 
foamy situations, one can define 
an operator that connects asymptotic density matrices~\cite{hawking}:
$\rho_{\rm out} \equiv {\rm Tr}_{M}| {\rm out} ><{\rm out} | 
= \$ ~\rho_{\rm in},~ \qquad \$ \ne S~S^\dagger, $ where 
the lack of factorization is attributed to the apparent loss of unitarity of the effective low-energy theory, defined as the part of the theory
accessible to low-energy observers who perform scattering experiments.
This defines what we mean by {\it particle phenomenology} in such  situations.
The \$ matrix is {\it not invertible}, and this reflects the 
effective unitarity loss. It is this property, actually, that leads to a 
{\it violation of CPT invariance}, at least in its strong form, in the 
sense that the generator of CPT symmetry is ill-defined. This has
been shown rigorously in ref.~\cite{wald}. 
However, as cautiously pointed out there, despite the 
strong violation of CPT in such a situation, one cannot exclude
the possibility that a {\it weak form of CPT invariance} 
remains, which is reflected in the possibility of the 
``experimentalist'' to prepare pure initial quantum states, $|\psi\rangle$, 
that evolve into pure asymptotic states, $|\phi\rangle$
(defining, in some sense, a ``decoherence-free subspace'' 
in the language of open systems), in such a way that 
CPT is preserved in the respective probabilities:
$P(\psi \to \phi ) = P(\theta ^{-1}\phi \to \theta \psi ) $,
where $\theta$: ${\cal H}_{\rm in} \to {\cal H}_{\rm out}$, 
with ${\cal H}$ denoting the appropriate Hilbert state spaces.
The notation is such that the CPT operator $\Theta$ acting on 
density matrices is: $\Theta \rho = \theta \rho \theta^\dagger, \quad  
\theta ^\dagger = - \theta^{-1} ~(\rm anti-unitary)$. 

As we shall discuss in this article, such issues can, in principle, 
be resolved experimentally, provided of course the sensitivity of 
the experiments is appropriate to the order expected from  
theoretical models of quantum gravity.

\subsection{Cosmological CPT Violation?} 

Another type of possible violation of CPT~\cite{mavromatosdecohe}, 
which falls within the 
remit of the theorem of \cite{wald}, may be associated with the recent 
experimental evidence from supernovae and temperature 
fluctuations of the cosmic microwave 
background on a current era acceleration of the Universe, and the fact that
more than 70 \% of its energy budget consists of Dark Energy.
If this energy substance is a positive 
cosmological constant (de Sitter Universe), $\Lambda > 0$, 
then there is a cosmic horizon, in the sense that in a flat 
Universe, as the data seem to indicate we live in, 
light emitted at some future moment $t_0$ in the cosmic time
takes an infinite time to cover the finite horizon radius.
The presence of a cosmic future horizon implies impossibility
of defining proper asymptotic states, and in particular
quantum decoherence of a matter quantum field theory in this de Sitter 
geometry, as a result of an environment of modes crossing the horizon. 
In view of \cite{wald}, then, this would imply
an analogous situation with the foam, discussed above, i.e. a
strong form of CPT violation. In this particular case, as the horizon is 
macroscopic, one would definitely have the evolution of an initial
pure quantum state to a mixed one, and probably no weak form of
CPT invariance would exist, 
in contrast to the black-hole case~\cite{mavromatosdecohe}.

\subsection{CPT Violation in the Hamiltonian, 
consistent with closed system quantum mechanics}

Another reason for CPT violation (CPTV) 
in quantum gravity is {\it spontaneous breaking of Lorentz symmetry} (LV), 
without necessarily 
implying decoherence.
This has also been argued to occur in string theory and other
models of quantum gravity~\cite{sme,kostelecky}, but, 
in my opinion, no concrete microscopic
model has been given as yet. In string theory, for instance, 
where a LV vacuum has been argued to exist~\cite{sme},
it was not demonstrated
that this vacuum is the energetically preferred one.
So far, most of the literature in this respect is 
concentrating on consistent parameterizations of extension of the 
standard model (SME)~\cite{sme}, which can be used to bound 
the relevant Lorentz and/or CPT violating parameters. 
An important difference of this approach, as compared with the decoherence one
is the fact that in SME or other spontaneous Lorentz symmetry breaking
approaches, the CPT Violation is linked merely with non-commutativity 
of the CPT operator, which otherwise is a well-defined quantum mechanical
operator, with the Hamiltonian of the system. In such models 
there is a well defined scattering matrix, and the usual phenomenology applies.
As we shall argue below, experimentally one can in principle disentangle
these two types of violation, by means of appropriate observables and their
time evolution properties.

\subsection{Order of Magnitude Estimates of Quantum Gravity Effects} 

At first sight, the CPT violating effects can be estimated to be 
strongly suppressed.
Indeed, naively, Quantum Gravity (QG) has a dimensionful coupling 
constant:
$G_N \sim 1/M_P^2$, where $M_P =10^{19}$ GeV is the Planck scale. 
Hence, CPT violating 
and decoherening 
effects may be expected to be suppressed
by terms (with dimensions of energy) of order 
$E^3/M_P^2 $, where $E$ 
is a typical energy scale of the low-energy 
probe. 
If such is the case, the current facilities seem far 
from reaching this sensitivities. But,  
as we shall mention below, high energy cosmic neutrino 
observations  
may well reach such sensitivities in the future. 
However, there may be cases where loop resummation and other 
effects 
in theoretical 
models 
may result in much larger CPT-violating effects of order: 
$\frac{E^2}{M_P}$.
This happens, for instance, in some loop gravity approaches to 
Quantum Gravity, or some non-equilibrium stringy models of space-time
foam involving 
open string excitations~\cite{mavromatosdecohe}. Such large effects 
can lie within the sensitivities of 
current or immediate future experimental facilities
(terrestrial and astrophysical), and hence can be, or are already, 
excluded~\cite{reviews}.

\section{Phenomenology} 

The phenomenology of CPT Violation is complicated, as there is no
single figure of merit, and the relevant sensitivities are 
highly model dependent. Below I outline briefly, 
the main phenomenological searches for CPT violation, and the respective
sensitivities, in the various major 
approaches to CPT Violation outlined above.For further details 
see refs.~\cite{sme,kostelecky,mavromatosdecohe,
reviews}.

\subsection{CPT Violation in the Hamiltonian} 

The main activities in this area concern: 
(i) Lorentz violation 
in extensions of the standard model~\cite{sme,kostelecky},
which provide an exhaustive phenomenological study of 
various Lorentz and CPT Violating effects in a plethora
of atomic, nuclear and particle (neutrinos) physics experiments.
Modified Dirac equation in the presence of external 
gauge fields can be used for constraining CPTV and Lorentz 
violating parameters by means 
of, say, antimatter factories~\cite{factories}.
I will not discuss further such tests here. For details I refer
the reader to the literature~\cite{sme,kostelecky,mavro_leap03}.
Specific tests for CPTV in the antihydrogen system, via precision
measurements of its hyperfine structure, 
can be found in \cite{hayano}. At present, the most sensitive
of the parameters of CPT and Lorentz violation, $b$ in the notation
of \cite{kostelecky} 
can be constrained to be smaller than $b < 10^{-27}$ GeV 
(or $b < 10^{-31}$ GeV in masers). Since there is 
no microscopic model underlying the standard model extension at present
it is difficult to interpret such small bounds, as far as sensitivity 
at Planckian energy scales is concerned. In the naive dimensional estimate
that $[b]=E^2/M_{QG}$, where $M_{QG}$ is a quantum gravity scale, 
one obtains sensitivities to scales up to $M_{QG}= 10^{27}$ GeV, thereby
tending to excluding linear suppression by the Planck mass. 
However, if, as expected in such models, there are quadratic (or higher)
suppressions by $M_P$, then we are some 10 orders of magnitude away from 
Planck scale physics at present.

(ii) Tests of modified dispersion relations for matter probes.
Stringent tests for charged fermions (e.g. electrons) are 
provided by synchrotron radiation measurements from 
astrophysical sources, e.g. Crab nebula~\cite{crab,reviews},
whilst the most accurate tests of modified dispersion relations for photons
are at present provided by Gama-Ray-Burst observations on arrival times
of radiation at various energy channels~\cite{nature}. 
The present observations of photons from Gamma Ray Bursts 
and Active Galactic Nuclei (AGN) concern energy scales up to a few MeV at most.
If the quantum gravity effects are then linearly suppressed 
by the quantum gravity scale, then the sensitivity is $M_{QG} > 10^{16}$ GeV,
otherwise is much less. There is an issue here concerning the universality of 
quantum gravity effects. As argued in \cite{equiv},
studies in certain microscopic models of modified dispersion
relations in string theory, have indicated that only photons
and standard model gauge bosons
may be susceptible to quantum gravity effects in some models.
This is due to some stringy gauge symmetry protection
of chiral matter particles against such effects, to all orders
in perturbation theory in a low-energy quantum field theory context. 
It is in the above sense that it is important to obtain
limits on such effects from various sources and for various systems.

(iii) As a final comment in this type of CPT violation we also mention the 
possibility that quantum gravity may induce {\it non hermiticity} 
of the effective low energy Hamiltonian, in the sense of 
{\it complex} coupling constants appearing in low energy theories~\cite{okun}. 
This may lead to interesting phenomena, for instance 
complex anomalous magnetic moments of, say, protons,
from which one may place stringent constraints in such effects. 
Since in all microscopic models we have in our disposition so far
such imaginary effects do not appear we shall not pursue 
this discussion further inhere. However, we stress again, 
we cannot exclude this possibility from appearing in future 
microscopic models of 
quantum gravity.

\subsection{Quantum Gravity Decoherence and CPT Violation} 

\emph{Neutral Mesons (Single Kaon states)} 

Quantum Gravity  may induce decoherence and oscillations 
among neutral mesons, such as kaons 
$K^0 \to {\overline K}^0$~\cite{ehns,lopez}.  
The modified evolution equation for the respective density matrices
of neutral kaon matter can be parametrized as follows~\cite{ehns}: 
$$\partial_t \rho = i[\rho, H] + \delta\H \rho~,$$ 
where $H$ is the standard Kaon Hamiltonian, 
and the ``entanglement'' matrix is given by
$$ {\delta\H}_{\alpha\beta} =\left( \begin{array}{cccc}
 0  &  0 & 0 & 0 \\
 0  &  0 & 0 & 0 \\
 0  &  0 & -2\alpha  & -2\beta \\
 0  &  0 & -2\beta & -2\gamma \end{array}\right)~.$$
Positivity of $\rho$ requires:
$\alpha, \gamma  > 0,\qquad \alpha\gamma>\beta^2$.
Notice that 
$\alpha,\beta,\gamma$ violate CPT in the sense of 
an induced {\it microscopic time irreversibility} of \cite{wald}, 
as being associated with decoherence, but also they violate CP
since they do not commute
with the CP operator~\cite{lopez}:  
$CP = \sigma_3 \cos\theta + \sigma_2 \sin\theta$,$~~~~~[\delta\H_{\alpha\beta}, CP ] \ne 0$.

An important remark is now in order. 
We should distinguish two types of CPTV:
(i) CPTV within Quantum Mechanics:
$\delta M= m_{K^0} - m_{{\overline K}^0}$, $\delta \Gamma = \Gamma_{K^0}-
\Gamma_{{\overline K}^0} $. 
This could be  due to (spontaneous) Lorentz violation (c.f. above).\\
(ii) CPTV through decoherence $\alpha,\beta,\gamma$ 
(entanglement with QG `environment', leading to modified 
evolution for $\rho$ and  $\$ \ne S~S^\dagger $). 

The important point is that the two types of CPTV can be {\it disentangled 
experimentally}~\cite{lopez}. 
Experimentally, the best available bounds of $\alpha,\beta,\gamma$ 
parameters to date 
for single neutral Kaon states come from 
CPLEAR measurements~\cite{cplear} 
$\alpha < 4.0 \times 10^{-17} ~{\rm GeV}~, ~|\beta | < 2.3. \times
10^{-19} ~{\rm GeV}~, ~\gamma < 3.7 \times 10^{-21} ~{\rm GeV} $,
which are not much different from 
theoretically expected values in the most optimistic of the scenaria,
involving linear Planck-scale suppression 
$\alpha~,\beta~,\gamma = O(\xi \frac{E^2}{M_{P}})$.
For more details we refer the reader to the 
literature~\cite{lopez,mavro_leap03}.

\emph{Neutral Meson factories: entangled meson states}

The above-described decoherence formalism can be used to 
derive the (non-unitary) evolution of the entangled products of the $\phi$ or 
$\Upsilon$ decays in meson factories. The requirement of 
\emph{complete positivity} of the respective
density matrices imposes further restrictions 
among the decoherence parameters in that case, 
amounting to setting $\beta = 0$, $\alpha = \gamma \ne 0$~\cite{benatti}.

An entirely novel observable, 
exclusive to a breakdown of CPT Violation through 
decoherence, in which case the CPT operator is not well defined, 
pertains to observations of the modifications of 
Einstein-Podolsky-Rosen entangled states
of neutral mesons in meson factories ($\phi-$ or B-factories)~\cite{bernabeu}.
These modifications concern the nature of the products of the decay 
of the neutral mesons in a factory on the two sides of the detector.
For instance, for neutral kaons, if the CPT operator is a well defined 
operator, even if it does not commute with the Hamiltonian of the system,
the products of the decay contain states $K_LK_S$ only.
On the contrary, in the case of CPT breakdown 
through decoherence (ill defined CPT operator), one obtains
in the final state, 
{\it in addition} to $K_LK_S$,
also 
$K_SK_S$ and/or $K_LK_L$ states. 
In a similar manner, 
when this formalism of CPT breaking through decoherence 
is applied to B-systems, one observes that flavor tagging fails 
in B-factories as a result of such CPTV modifications~\cite{bernabeu}.

\emph{Neutrinos} 

Similarly to neutral mesons, one could have decoherening modifications 
of the oscillation probabilities for neutrinos. 
Due to lack of space we shall not describe the details here.
We refer the interested reader in the literature~\cite{mavromatosdecohe}.
We only mention that at present there is no experimental evidence
in neutrino physics on quantum gravity decoherence effects, only 
very stringent bounds exist.
Although, at first sight, it appears that neutrino anomalies,
such as the LSND result~\cite{lsnd} indicating asymmetric rates for 
antineutrino oscillations as compared to 
neutrino ones, can be fit in the context of a 
decoherence model with mixed energy 
dependences, $E$ and $1/E$, 
and with different orders
of the decoherence parameters between particle and antiparticle 
sectors~\cite{barenboim}, 
nevertheless recent spectral distortions from KamLand~\cite{kamland}
indicate that standard oscillations are capable of explaining these 
distortions, 
excluding the order of decoherence in the antineutrino sector 
necessary to match the LSND result with the rest of the available 
neutrino data.
This leaves us with the following bounds as far as decoherence 
coefficients $\gamma$ are concerned for neutrinos~\cite{lizzi}:
in a parametrization  
$\gamma \sim \gamma_0 (\frac{E}{\rm GeV})^n$, with $n=0,2,-1$,
we have: (a) $n=0$, $\gamma_0 < 3.5 \times 10^{-23}$ GeV,
(b) $n=2$, $\gamma_0 < 0.9 \times 10^{-27}$ GeV (compare with the 
neutral Kaon case above), and (c) $n=-1$, $\gamma_0 < 2 \times 10^{-21}$ GeV.
Very stringent limits on quantum gravity decoherence 
may be placed from future observations of high-energy 
neutrinos 
from extragalactic sources~\cite{mavromatosdecohe}
(Supernovae, AGN), if, for instance, Quantum Gravity 
induces lepton number violation and/or flavor
oscillations. 
We therefore see that neutrinos provide the most sensitive probe at present
for tests of 
quantum gravity decoherence effects, provided the latter 
pertain to these particles.

\emph{Disentangling Matter Effects from Genuine Quantum Gravity 
induced Decoherence 
effects} 

However, there is an important aspect in all such decoherence searches,
which should not be overlooked. This concerns ``fake'' decoherence 
effects as a result of the passage of the particle probe, e.g. neutrino,
through matter media. Due to the apparent (``extrinsic'') breaking of 
CPT in such a case, one obtains fake violations of the symmetry,
which have nothing to do with genuine microscopic gravity effects.
Matter effects, for instance, include standard CPTV differences in neutrino 
oscillation probabilities of the form: 
$P_{\nu_\alpha\to\nu_\beta} \ne P_{\bar \nu_\beta \to \bar \nu_\alpha}$,
non-zero result for the so-called 
$A_{CPT}^r$ asymmetry to leading order, 
in the case of Kaons through 
a regenerator~\cite{lopez}, as well as uncertainties in the energy 
and oscillation length of neutrino 
beams, which result in damping factors in front of terms in the 
respective oscillation probabilities. Such factors  
are of similar nature to those induced by 
decoherence $\gamma$ coefficients~\cite{olsson}. 
Nevertheless, the energy and oscillation length dependence of the 
effects is different~\cite{mavromatosdecohe,barenboim}, 
and allows disentanglement from 
genuine effects, if the latter are present and are of comparable order
to the matter effects. 
Namely, quantum gravity effects are in general
expected to increase with the energy of the neutrino probe,
as a result of the fact that the higher the energy the stronger the 
back reaction onto the space time. This should be contrasted with
ordinary matter effects, which are expected to decrease with the energy
of the matter probe. Also, the damping factors in the case of 
neutrinos with uncertainty in their energy have a different 
oscillation length dependence, as compared with quantum gravity 
decoherence effects.
Note that matter decoherence effects may 
be tiny~\cite{olsson},  
with the 
appropriate coefficients of order smaller than $\gamma_{\rm fake} < 
10^{-24}$ GeV, thereby leading to the temptation of identifying 
(incorrectly) such effects with genuine microscopic effects.

\section{Conclusions} 

There are various ways by means of which 
Quantum Gravity (QG)-induced CPT breaking can occur, 
which are in principle independent of each other. 
For instance, quantum decoherence
and Lorentz Violation are in general 
independent effects. There is no single figure of merit of CPT violation,
and the associated sensitivity depends on the way CPT is broken, and 
on the relevant observable. The pertinent phenomenology is not simple.

Neutrino (astro)physics may provide 
some of the most stringent (to date) constraints on QG CPT 
Violation~\cite{icecube}.
There are interesting theoretical issues on Quantum Gravity decoherence
and neutrinos, 
which go as far as the origin of 
neutrino mass differences (which could even be due to space time foam effects,
thus explaining their smallness),  
and their contributions to the Dark Energy~\cite{barenboim2}. 

Quantum-Gravity induced decoherence CPTV, in which 
the CPT generator is ill defined, 
may lead to interesting
novel observables with high sensitivity in the near future,
associated with 
entangled state modifications in 
neutral meson factories (B, $\phi$).

Thus, it seems that 
a century after {\it annus mirabilis}, there is still 
a long way to go before an 
understanding of Quantum Gravity is achieved.
But the challenge is there, and we think that there may be 
pleasant experimental surprises in the near future. This is 
why experimental searches of CPT violation and other quantum gravity 
effects along the lines presented here are worth pursuing.

\begin{theacknowledgments}
The author would like to thank Prof. W. Oelert and the other organizers
of the LEAP05 conference in Bonn/GSI for the invitation, and for creating
such a stimulating meeting. 
\end{theacknowledgments}

\end{document}